\documentclass[12pt]{iopart}

\usepackage{graphicx}
\graphicspath{{figures/}}

\bibstyle{unsrt}

\newcommand{\DeltaM}{\ensuremath{\Delta_{\rm M}}}
\newcommand{\DeltaA}{\ensuremath{\Delta_{\rm A}}}

\providecommand{\ket}[1]{\ensuremath{\left\vert #1\right\rangle}}
\providecommand{\bra}[1]{\ensuremath{\left\langle #1\right\vert}}

\providecommand{\comm}[2]{\left[#1,#2\right]}
\providecommand\avr[1]{\left\langle #1 \right\rangle}

\providecommand\real[1]{\ensuremath{\Re\left\{#1\right\}}}

\providecommand{\lp}{\left(}
\providecommand{\rp}{\right)}
\providecommand{\lbr}{\left\{}
\providecommand{\rbr}{\right\}}

\usepackage{color}

\begin{document}

\title{Optical bistability in strong-coupling cavity QED with a few atoms}
\footnote{This work was supported by the EU FP7 (ITN, CCQED-264666), the Hungarian National Office for Research and Technology under the contract ERC\_HU\_09 OPTOMECH, and the Hungarian Academy of Sciences  (Lend\"ulet Program, LP2011-016).}
\author{A.~Dombi, A. Vukics and P.~Domokos}
\address{Wigner Research Centre for Physics, Hungarian Academy of Sciences, H-1525 Budapest, P.O.~Box 49}
\eads{\mailto{dombi.andras@wigner.mta.hu}, \mailto{peter.domokos@wigner.mta.hu}}

\begin{abstract}
We present exact numerical solutions of the damped-driven Jaynes--Cummings model adapted to describe absorptive optical bistability in the limit of a few atoms strongly coupled to a high-finesse resonator. We show that the simplifying semiclassical result for many physical quantities of interest is well reproduced by the quantum model including even with only a few atoms in the strongly coupled system.  Nontrivial atom-field quantum cross-correlations show up in the strong-driving limit. 
\end{abstract}


\maketitle

\section{Introduction}

Optical bistability is an experimentally accessible and controllable example for a non-equilibrium phase transition in a damped-driven open system \cite{Lugiato1984II}. The bistability effect has been observed in various systems \cite{Gibbs1976Differential,Rempe1991Optical,Sauer2004Cavity,Elsasser2004Optical,Gupta2007Cavity,Brennecke2008Cavity,Vengalattore2008Optical}  in which there is a significant nonlinearity in the interaction between a radiation field and a polarizable medium. Interestingly, the required nonlinear coupling can be reached in different regimes of light--matter interaction ranging from the microscopic quantum to the semiclassical mean-field dominated one. Originally, the bistability has been studied in the transmitted power through a large volume resonator filled with a resonant atomic vapour as a macroscopic saturable absorber \cite{Gibbs1976Differential}. The development of microscopic cavity QED systems with strong coupling between atomic dipole and radiation field led to the observation of bistability in the input-output signal for a low number of atoms ($N<100$) \cite{Rempe1991Optical,Sauer2004Cavity}. Because of the stochastic distribution of the atoms within the cavity mode volume \cite{Carmichael1999Multiatom}, most of the quantum features were suppressed and the semiclassical theory  \cite{Agrawal1979Optical,Drummond1981Quantum,Carmichael1986Quantum} applies satisfactorily well to describe the observations even for such a small medium size. Although the coupling threshold for observing bistability can be in principle reached by a single atom, the fluctuations in the atomic trajectories due to optical forces hindered the systematic study of the quantum regime of bistability \cite{Hood1998RealTime} for long. Since then, the realization of controlled nonlinear coupling at the single atom few photons level has been achieved \cite{Schuster2008Nonlinear,Keller2003Deterministic,Kampschulte2010Optical}. In the strong coupling regime of cavity QED, the interplay of quantum fluctuations with nonlinear coupling at low intracavity photon number is expected to inherently modify  the optical bistability effect  \cite{Armen2006Lowlying}. Remarkably, the remnants of the semiclassical bistability have been recorded by means of a single atom coupled to the single mode of a high-finesse microresonator \cite{Kerckhoff2011Remnants}. Today, cavity QED allows for the controlled variation of the size of the atomic medium by single atom resolution \cite{Khudaverdyan2008Controlled,Reick2010Analyzing}. It is thus a suitable platform to explore the quantum corrections in a finite-size system to the semiclassical mean-field results in the vicinity of a critical point.  We aim at exploring the transition between the quantum and the semiclassical regimes of optical bistability in this paper. 

The Jaynes--Cummings model is fundamental to cavity QED  and describes the interaction between an single atomic dipole transition and a single mode of the radiation field sustained by a high-finesse resonator in the optical frequency domain. This model can straightforwardly extended to deal with few and many-atom systems. In the limit of large ensemble of independent atoms and weak  atom-mode coupling, the semiclassical Maxwell--Bloch equations usually adopted to describe optical bistability are rendered. The Jaynes--Cummings model is thus suitable to our computational study of the crossover regime in which the semiclassical solution gradually forms from the exact solution of an quantum model for increasing number of atoms.   

This paper is organized as follows: in \Sref{sec:System}, we present the physical system and the model, and the results in \Sref{sec:Results}. Finally, we conclude in \Sref{sec:Conclusion}.
     
\section{System and model}
\label{sec:System}

\begin{figure}
\begin{center}
\includegraphics[width=0.7\linewidth]{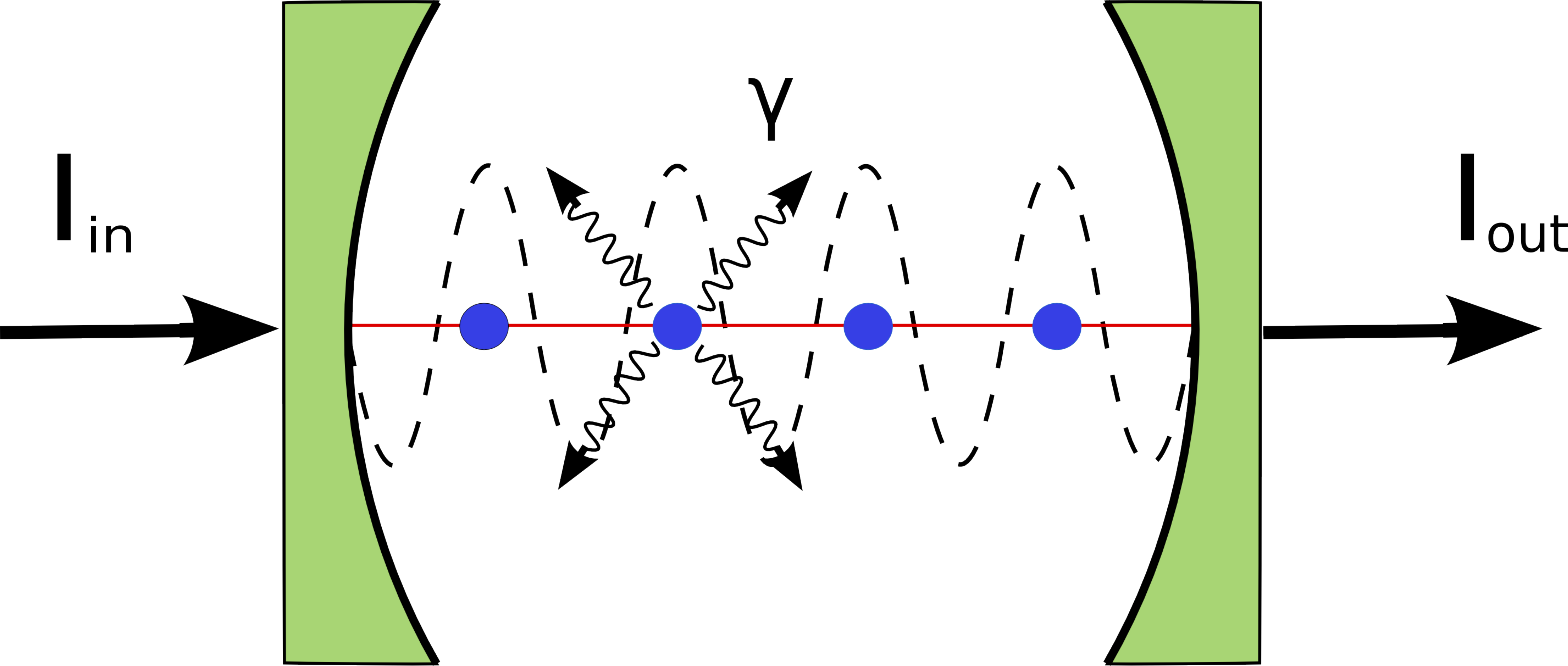}
\caption{Scheme of the cavity system coherently driven with intensity $I_{\rm in}$. The $N$ unmoving two-level atoms are identically coupled to a single mode of the cavity field. The atoms interact with independent reservoirs so that the spontaneous decay is an individual process, however, the corresponding damping rate is identical. The outcoupled field with intensity $I_{\rm out}$ is detected. Due to the nonlinearity of the saturable two-level atoms, the input--output function $I_{\rm in} \rightarrow I_{\rm out}$ may present multivalued regions, i.e., optical bistability and hysteresis.}
\label{fig:Scheme}
\end{center}
\end{figure}

We consider a fixed number ($N$) of identical two-level atoms with resonance frequency $\omega_A$ and linewidth (HWHM) $\gamma$ coupled to a single mode of a high-finesse cavity. The atoms are fixed at the antinodes of the field mode resulting in a uniform coupling to the atomic internal degrees of freedom with coupling strength $g$. The cavity is coherently driven with pump strength $\eta$ at a probe frequency $\omega$ detuned from the cavity mode by $\DeltaM=\omega-\omega_{\rm M}$ and from the atom by $\DeltaA=\omega-\omega_{\rm A}$. The cavity mode linewidth (HWHM) is denoted by $\kappa$.

In the electric-dipole and rotating-wave approximations the interaction between the single cavity mode and the atomic internal degrees of freedom is described by the Jaynes--Cummings Hamiltonian, which, in a frame rotating with the driving laser frequency $\omega$, reads ($\hbar=1$)
\begin{equation*}
H=-\DeltaM\,a^\dagger a - \DeltaA \sum_{i=1}^{N}\sigma^\dagger_i\,\sigma_i + i g \sum_{i=1}^{N} \lp a^\dagger\,\sigma_i-\sigma^\dagger_i\,a\rp + i \eta\lp a^\dagger-a\rp,
\end{equation*}
where $a$ and $a^\dagger$ are the bosonic annihilation and creation operators, while $\sigma_i$ and $\sigma^\dagger_i$ are lowering and raising operators for atom $i$. These latter complemented by the population inversion $\sigma_{z,i} = \sigma_i^\dagger \sigma_i - \frac{1}{2}$ and the unit operator form a complete set and their algebra is equivalent to that of the Pauli operators of a spin-$\frac{1}{2}$ particle. 

There are two dissipation channels of the system considered: the cavity-photon loss and the decay of the atomic excited states into the free-space modes of the electromagnetic field, which latter is considered for each atom separately. The reservoir is taken at zero temperature. The corresponding Master equation can be cast into Lindblad form with the quantum-jump operators $a$ and $\sigma_i$ for each $i$, respectively, to read
\begin{equation}
\fl\label{eq:Master}
\dot{\rho}=-i\comm{H}{\rho} + \kappa\lp 2 a \rho a^\dagger - a^\dagger a \rho - \rho a^\dagger a \rp + \gamma\sum_{i=1}^{N}\lp 2 \sigma_i \rho \sigma^\dagger_i - \sigma^\dagger_i\sigma_i \rho - \rho \sigma^\dagger_i\sigma_i  \rp.
\end{equation}
It is important to note that the two-level atoms decay independently. Therefore, although their coupling to the cavity mode is assumed to be symmetric, the ensemble of these spins cannot be replaced by a single collective spin. This is because the state of the system, via individual decays, leaves that subspace of the Hilbert space which corresponds to states of indistinguishable atoms, be they either fermions or bosons. Put otherwise, the individual decay allows for distinguishing the atoms.
 
\subsection{Full quantum solution}

Our primary method for solving the Master equation  (\ref{eq:Master}) consists in unravelling it into a set of Monte Carlo wave-function trajectories, whose ensemble average reproduces the density operator. Aiming at only the steady state, however, we can invoke the ergodic hypothesis and replace the ensemble averaging by time averaging over a single trajectory run for a very long time (much longer than the inverse of the smallest frequency of the system) \cite{VukicsQuantumCooling}. That is, instead of the numerically too demanding ensemble averaged
\begin{equation*}
\rho_{\rm ss}^{\rm ensemble}=\lim_{t\to\infty}\lim_{N_{\rm traj}\to\infty}\frac1{N_{\rm traj}}\sum_{n=1}^{N_{\rm traj}}\ket{\Psi_n(t)}\bra{\Psi_n(t)},
\end{equation*}
we consider the time averaged
\begin{equation*}
\rho_{\rm ss}^{\rm time}=\lim_{M\to\infty}\frac1{M}\sum_{m=1}^M\ket{\Psi(m\,\Delta t)}\bra{\Psi(m\,\Delta t)},
\end{equation*}
with an appropriately chosen \(\Delta t\) larger than the relaxation time of the system.

This method provides us with full information about the quantum steady state of the system, which is a substantial amount of data even for a moderate number of atoms which amount grows exponentially with $N$: The total dimension of the system is given by
\begin{equation*}
D_{\rm total}=2^ND_{\rm M}
\end{equation*}
yielding $5\cdot10^4$ for $N=8$ atoms and a generic value of $D_{\rm M}=200$ Fock states for the mode that we use. The actual simulations were performed using the C++QED framework \cite{VukicsCppQEDa,VukicsCppQEDb}. In this way, we are able to treat $N=2\dots8$ atoms and to monitor quantum statistical properties along the upper branch of the bistability curve, which goes well beyond the scope of previous attempts \cite{Martini1993Optical,Clemens2000Nonclassical}. In particular, we will see that this range is sufficient to explore the emergence of the semiclassical bistability behaviour.

In the following, we will also use the steady-state reduced density operator of the cavity mode:
\begin{equation*}
\rho_{\rm ss:mode}=\Tr_{\rm atoms}\lbr\rho_{\rm ss}\rbr
\end{equation*}

\subsection{Semiclassical limit}
The results stemming from the full quantum simulations will be compared to results from a mean-field approximation \cite{Armen2006Lowlying}. To this effect, we first derive a set of Heisenberg--Langevin equations equivalent to \Eref{eq:Master}:
\begin{eqnarray}
\dot{a}&=&\lp i\DeltaM-\kappa\rp a + g\sum_{i=1}^N\sigma_i + \eta +\xi, \nonumber \\
\dot\sigma_i&=&\lp i\DeltaA-\gamma\rp\sigma_i + 2 g\,\sigma_{z,i}\,a + \zeta_i, \nonumber \\
\dot\sigma_{z,i}&=&-\gamma\lp\sigma_{z,i}+\frac12\rp - g \lp\sigma^\dagger_i\,a + a^\dagger\,\sigma_i\rp + \zeta_{z,i} . \label{eq:HeisenbergLangevin}
\end{eqnarray}
The last term in each equations above represents the quantum noise which is defined by diffusion coefficients
\begin{eqnarray}
\avr{\xi(t_1)\,\xi^\dag(t_2)} &=& 2\kappa\,\delta(t_1-t_2),\nonumber \\
\avr{\zeta_i(t_1)\,\zeta_j^\dag(t_2)} &=& 2\gamma\,\delta_{i,j}\,\delta(t_1-t_2), \nonumber\\
\avr{\zeta_{z,i}(t_1)\,\zeta_{z,j}(t_2)} &=& 2\gamma\lp\avr{\sigma_{z,i}}+\frac 12\rp\delta_{i,j}\,\delta(t_1-t_2), \nonumber\\
\avr{\zeta_{z,i}(t_1)\,\zeta_j^\dag(t_2)} &=& 2\gamma\avr{\sigma_i^\dag}\delta_{i,j}\,\delta(t_1-t_2).
\end{eqnarray}
All other correlation functions vanish. 

The symmetric coupling of the atoms to the cavity mode suggests that we introduce the total spin operators $\Sigma=\sum_{i=1}^N\sigma_i$, $\Sigma_z=\sum_{i=1}^N\sigma_{z,i}$. These operators do not constitute a complete operator set in the atomic Hilbert space. However, one can obtain a closed set of equations by summing over $i$ in Eqs.~(\ref{eq:HeisenbergLangevin}):
\begin{eqnarray}
\dot a &=&\lp i\DeltaM-\kappa\rp a + g\,\Sigma + \eta +\xi, \nonumber\\
\dot\Sigma &=&\lp i\DeltaA-\gamma\rp\Sigma + 2 g\,\Sigma_z\,a + N\,\Xi ,\nonumber\\
\dot\Sigma_z &=&-\gamma\lp\Sigma_{z}+\frac{N}{2}\rp - g\lp\Sigma^\dagger\,a + a^\dagger\,\Sigma\rp + N\,\Xi_{z}, \label{eqs:HeisenbergLangevinCollective}
\end{eqnarray}
If the atoms were inhomogeneously coupled to the mode, i.e., by a constant coupling strength $g_i$ different for each atom $i=1\ldots N$, the resulting system of equations would not be closed. This was the main motivation behind our choice of uniform coupling: in this way the size of the atomic ensemble can be easily accounted for via the single parameter $N$ without increasing the number of parameters. The semiclassical limit can be derived by splitting the cavity mode amplitude and the collective spin variables to mean-field and quantum-fluctuation components. On introducing the scaled mean field and fluctuation variables, i.e., $a= \sqrt{N} (\alpha+\delta a)$, $\Sigma=N (S+\delta \Sigma$), and $\Sigma_z=N (S_z+\delta \Sigma_z)$,  the c-numbers $\alpha=\avr a /\sqrt{N}$, $S=\avr\Sigma /N$, and $S_z=\avr{\Sigma_z}/N$ representing the mean field, obey the well-known Maxwell--Bloch equations:
\begin{eqnarray}
\dot\alpha &=&\lp i\DeltaM-\kappa\rp\alpha + \sqrt{N} g\, S + \frac{\eta}{\sqrt{N}}, \nonumber\\
\dot S &=&\lp i\DeltaA-\gamma\rp S + 2 \sqrt{N} g\,S_z\,\alpha, \nonumber\\
\dot S_z &=&-\gamma\lp S_{z}+\frac{1}{2}\rp - \sqrt{N}g\lp S^*\,\alpha + \alpha^*\,S\rp. \label{eq:MeanField}
\end{eqnarray}
The dynamical equations for the fluctuations are linearized, 
\begin{eqnarray}
\dot{\delta a} &=&\lp i\DeltaM-\kappa\rp \delta a + \sqrt{N}g\,\delta\Sigma + \frac{\xi}{\sqrt{N}}, \nonumber\\
\dot{\delta\Sigma} &=&\lp i\DeltaA-\gamma\rp\delta\Sigma + 2\sqrt{N} g\lp S_z\,\delta a+\alpha\,\delta\Sigma_z\rp + \Xi, \nonumber\\
\dot{\delta\Sigma_z} &=&-\gamma\,\delta\Sigma_{z}-2\sqrt{N}g\,\real{S^*\,\delta a+\alpha\,\delta\Sigma^\dagger}+ \Xi_{z}, \label{eqs:Linearized}
\end{eqnarray}
and their driving terms arise from the quantum fluctuations associated with the dissipative processes, i.e., $\Xi= \frac{1}{N} \sum_{i=1}^N\zeta_i$ and $\Xi_z= \frac{1}{N} \sum_{i=1}^N\zeta_{z,i}$. The non-vanishing diffusion coefficients read
\begin{eqnarray}
\avr{\Xi (t_1) \Xi_z (t_2)} = \frac{2 \gamma}{N} \, S \,  \delta(t_1-t_2),\nonumber\\
\avr{\Xi_z (t_1) \Xi_z(t_2)}= \frac{2 \gamma}{N} \, \lp S_z +\frac{1}{2}\rp \,  \delta(t_1-t_2),\nonumber\\
\avr{\Xi (t_1) \Xi^\dagger(t_2)} = \frac{2 \gamma}{N} \,  \delta(t_1-t_2) \,.
\end{eqnarray}
Note that in the bistability regime, for the linearization procedure of this semiclassical calculation, one must select one of the mean field solutions to insert in $S$ and $S_z$ above.


It is important to notice that the mean-field equations are invariant under the variation of the atom number $N$ provided the coupling $g$ and the driving amplitude $\eta$ are simultaneously scaled such that $N g^2$ and $\eta/\sqrt{N}$ are kept constant. At the same time, the quantum fluctuations are reduced with increasing atom number as can be directly seen from the fact that the diffusion coefficients are proportional to $\gamma/N$. Therefore, one can expect that the general solution must tend to that of the mean-field equations in the large-$N$ limit.  However, the mean-field approximation neglects the consequences of the nonlinear term $\Sigma_z\,a $ and alike, appearing in the operator equations (\ref{eq:HeisenbergLangevin}). As we will see in what follows, the full quantum calculation leads to significant quantum correlations that can be attributed to this very term.

\section{Scaling of optical bistability with the atom number}
\label{sec:Results}
We now analyse the dependence of various steady-state characteristics of the bistable atom-cavity system on the atom number $N$. In order to ensure the invariance of the mean-field equations (\ref{eq:MeanField}) under changes of $N$, the coupling coefficient $g$ and the driving amplitude $\eta$ is scaled in such a way that the cavity cooperativity parameter $C=N\,g^2/(2 \kappa\gamma)$ and $\eta/\sqrt{N}$ remain constant. Any significant variation of measurable quantities as a function of $N$ thus reveals the contribution of non-trivial quantum correlations neglected in the semiclassical approach \cite{Carmichael1986Quantum}. 

\begin{figure}
\begin{center}
\includegraphics[width=0.9\linewidth]{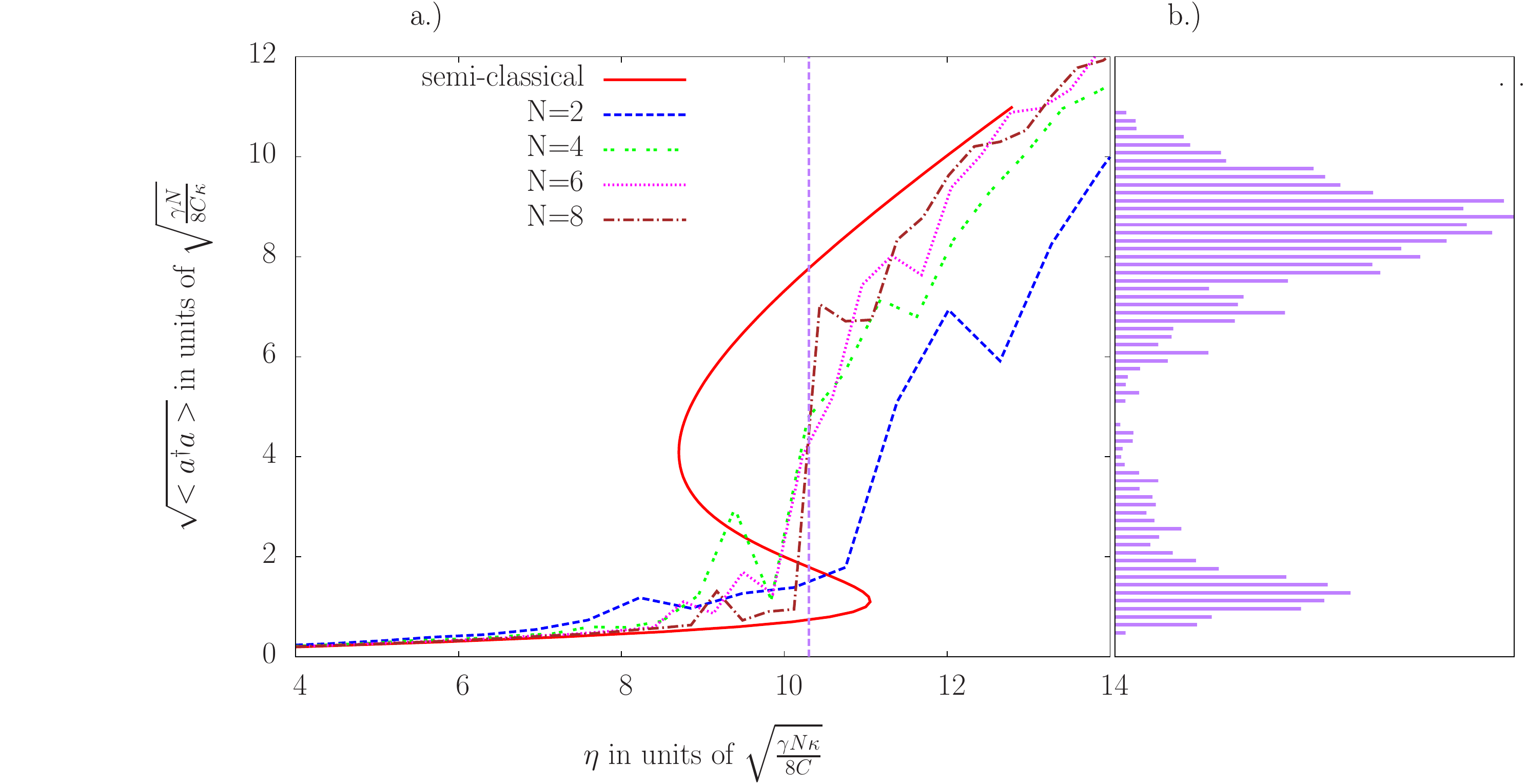}
\caption{(a) Outcoupled amplitude in steady state as a function of the drive amplitude calculated for $N=2,\,4,\,6,\,8$ @ $\Delta_A=\Delta_C=0$, $\kappa=\gamma/2$, and cooperativity $C=10$. For comparison, the semiclassical (mean-field) result is plotted in solid red line with the bistability signalled by the S-shape of the curve. (b) Histogram (in arbitrary units) of the outcoupled amplitude for $N=8$ at the driving marked by the vertical dashed line in (a).}
\label{fig:steadyStateIntensity}
\end{center}
\end{figure}

The semiclassical benchmark of optical bistability is the S-shaped curve of the output amplitude as a function of the input amplitude ($\eta$), as displayed as the solid red curve in \Fref{fig:steadyStateIntensity}, calculated from the mean-field equations (\ref{eq:MeanField}). The output and input field amplitudes are the square roots of the respective intensities. In the quantum simulation yielding the steady-state density operator, the closest quantity comparable thereto is the square root of the steady-state quantum average of the mode photon number: $\sqrt{\avr{a^\dagger a}}$, the output amplitude reading
\begin{equation*}
A_{\rm out}=\sqrt{\kappa\avr{a^\dagger a}}\quad \lp=\sqrt{I_{\rm out}}\rp.
\end{equation*}

Obviously, a bistability behaviour cannot manifest in this quantity alone, as it must remain single-valued for arbitrary driving strength. Accordingly, in \Fref{fig:steadyStateIntensity}(a) we see that the curves from the quantum calculation converge to the semiclassical curve only outside the bistability region. For $N=8$, the convergence is already quite close. In the semiclassically predicted bistability regime, the quantum curves are characterised by substantial noise. This noise is intrinsic to the quantum system and is related to the semiclassical bistability, because in this regime the quantum average of the photon number fluctuates between the possible values represented by the branches of the bistability curve. (Upon much longer time averaging, these curves would certainly smoothen, and the quantum average would assume a value corresponding to the average of the possible values.) From this Figure, it is also apparent that the above-mentioned scaling as a function of $N$ that we expect from the mean-field equations is indeed correct because it provides for an overlap of the curves taken for different numbers, see the results for $N=6$ and $N=8$.

\begin{figure}
\begin{center}
\newlength{\widthWith}\setlength{\widthWith}{.45\linewidth}
\newlength{\heightWithout}\settoheight{\heightWithout}{\includegraphics[width=\widthWith]{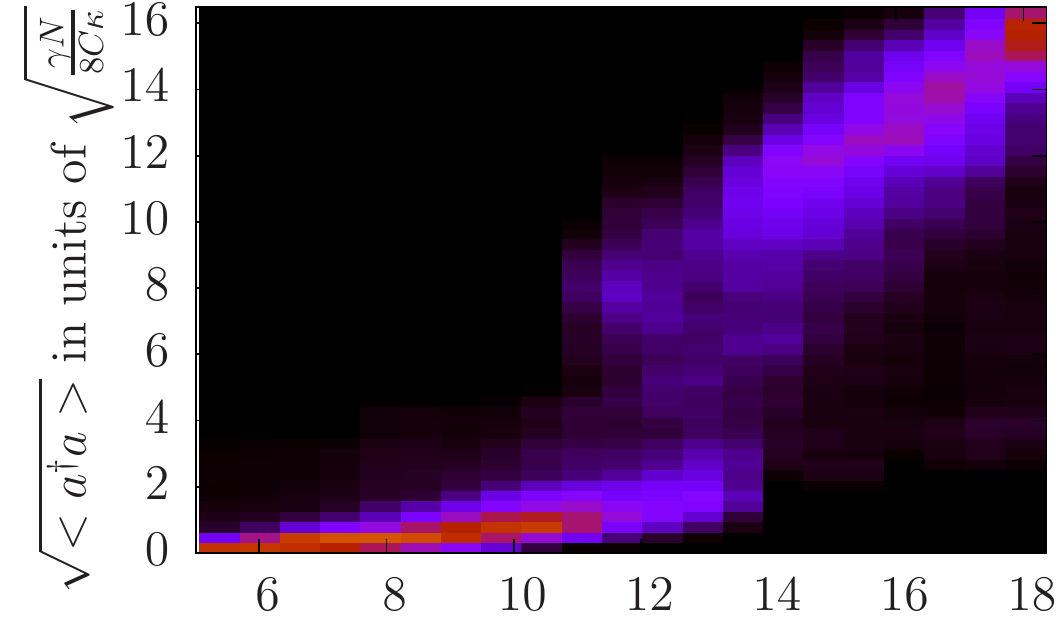}}
\newlength{\widthWithout}\settowidth{\widthWithout}{\includegraphics[height=\heightWithout]{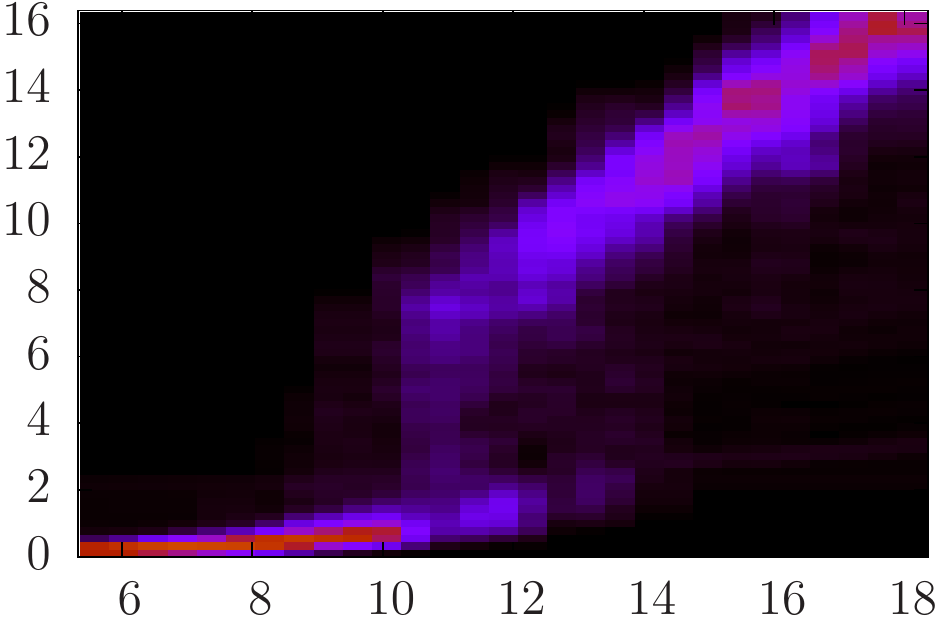}}
\newlength{\heightWith}\settoheight{\heightWith}{\includegraphics[width=\widthWith]{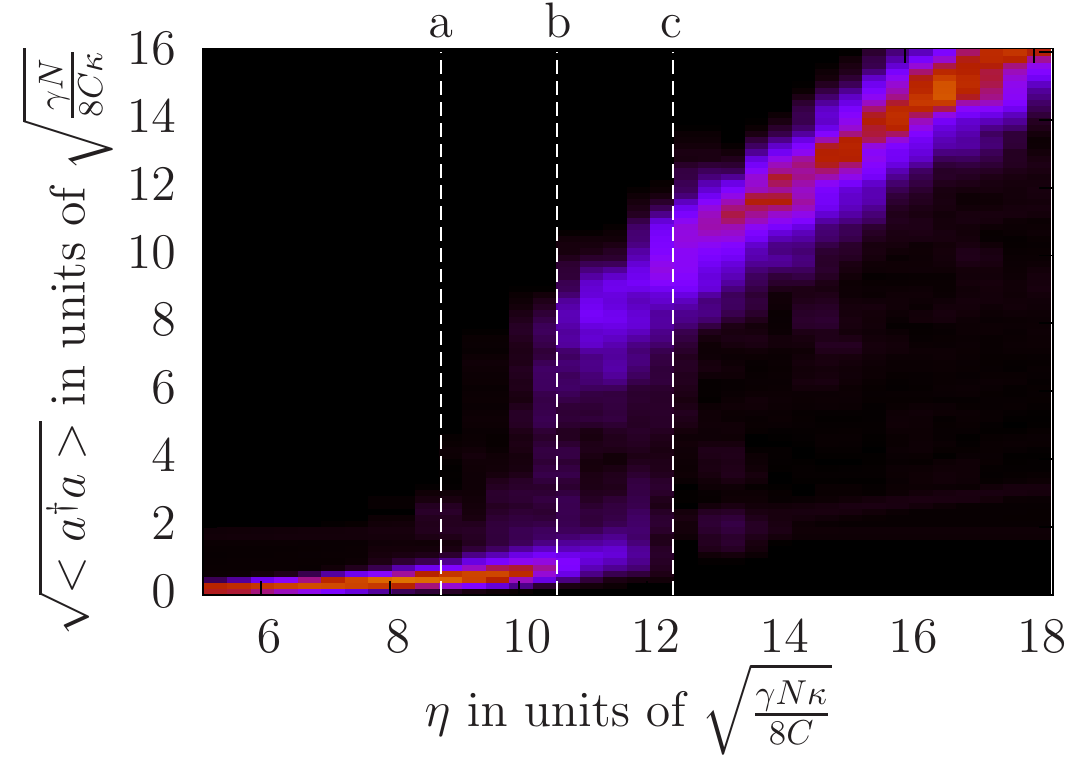}}
\begin{tabular}{l l}
$N=2$ & $N=4$\\
\raisebox{-0.02\heightWithout}{\includegraphics[width=\widthWith,height=1.02\heightWithout]{fig3a}} & \includegraphics[height=\heightWithout]{fig3b}\\
$N=6$ & $N=8$\\
\includegraphics[width=\widthWith,height=1.05\heightWith]{fig3c} & \includegraphics[width=\widthWithout]{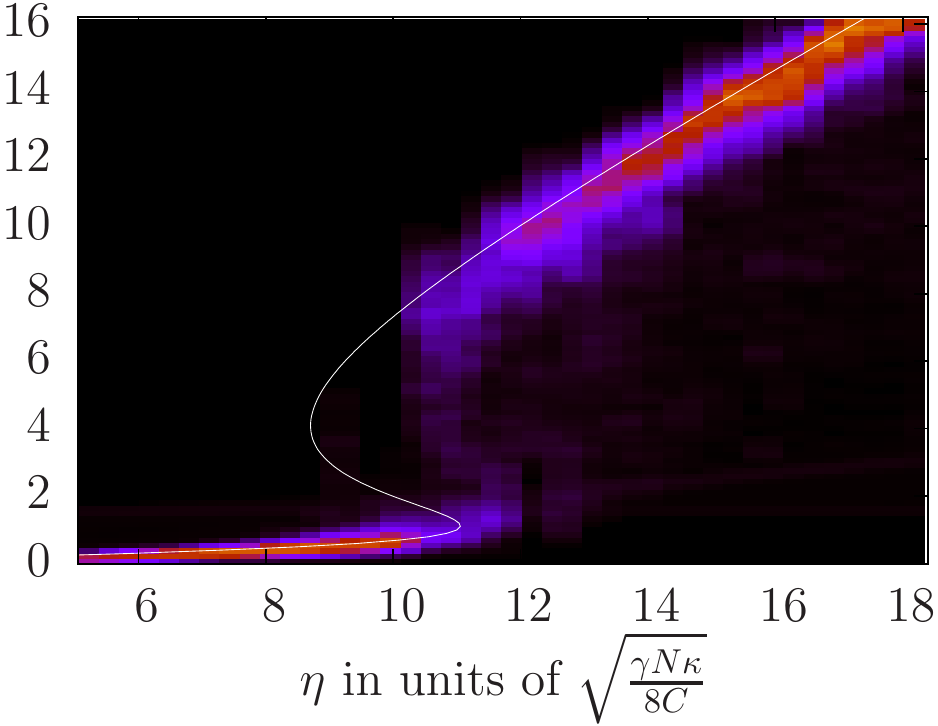}
\end{tabular}
\end{center}
\caption{Outcoupled-amplitude histograms as functions of the input amplitude registered in the same way as in \Fref{fig:steadyStateIntensity}(b). The histograms are now plotted in colour-code (arbitrary units). For $N=8$, the semiclassical (mean-field) bistability curve is displayed in white. Same parameters as in \Fref{fig:steadyStateIntensity}.}
\label{fig:histogram}
\end{figure}

The correspondence between the semiclassical and the quantum results in the bistability region can be best seized by a histogram of the outcoupled field amplitude as displayed in \Fref{fig:steadyStateIntensity}(b). The histogram is registered along the same single quantum trajectory that is used for time averaging. This plot clearly manifests that the photon-number distribution, and, accordingly, the outcoupled amplitude is a bimodal distribution in the semiclassical bistability range. We further elaborate on this concept in \Fref{fig:histogram}, where the $A_{\rm out}$ histograms are displayed as a function of the input amplitude. For $N=8$, the convergence to the semiclassical curve is quite close also in the bistability region.

Let us note that qualitatively identical, though much more coarse-grained results can be obtained using the steady-state photon-number distribution calculated from the steady-state density matrix reduced to the mode. In this case, the diagonal of this density matrix replaces the histogram taken along a single trajectory.

\subsection{Quantum statistics of the light field}

Many quantum statistical properties of the cavity field mode, including first-order quadrature correlations can be visualised and most conveniently discussed in terms of the Wigner function. For a given mode density operator expressed in Fock basis, this reads
\begin{eqnarray}\nonumber\fl
W[\rho](x,y)=&\frac{2\,e^{-2\lp x^2+y^2\rp}}{\pi}\sum_{m,n}\frac{\rho_{m,n}}{\sqrt{m!n!}}(-1)^n(2i)^{-m-n}
\\&\times\sum_{k'=0}^m\sum_{k''=0}^n
\lp\begin{array}{c}m\\k'\end{array}\rp
\lp\begin{array}{c}n\\k''\end{array}\rp
i^{k'+k''}(-1)^{k'}\,H_{k'+k''}(-2x)\,H_{m+n-k'-k''}(2y).
\label{eq:WignerFunction}
\end{eqnarray}

\begin{figure}
\begin{center}
\begin{tabular}{l l}
(a)&(b)\\
\includegraphics[width=.4\linewidth]{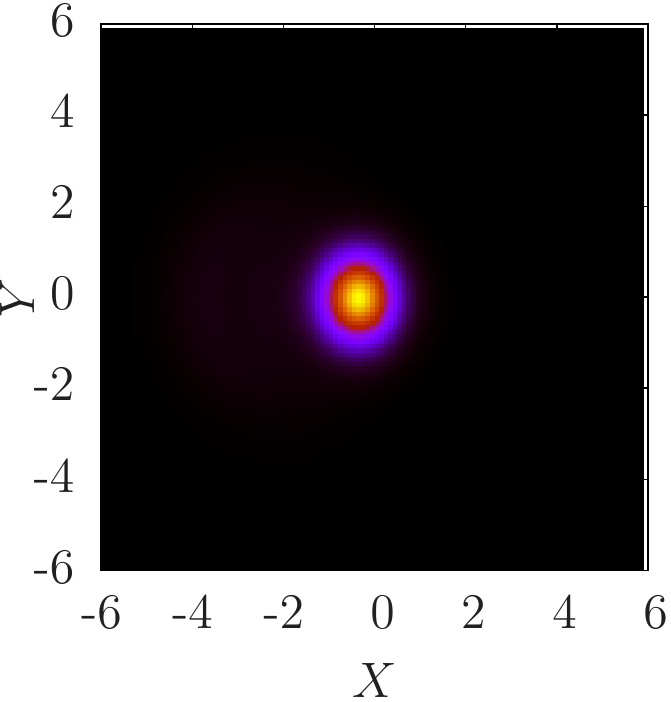} & \includegraphics[width=.4\linewidth]{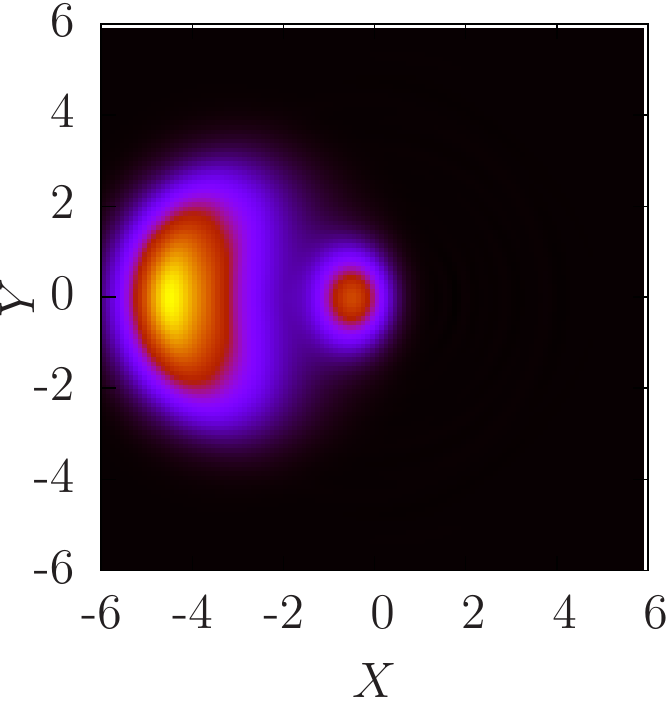}\\
(c)&(d)\\
\includegraphics[width=.4\linewidth]{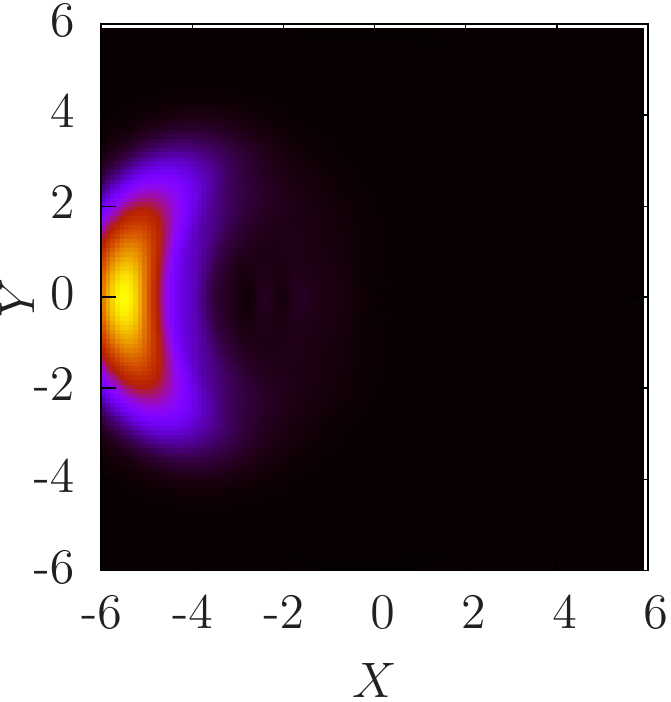} & \includegraphics[width=.4\linewidth,height=.4\linewidth]{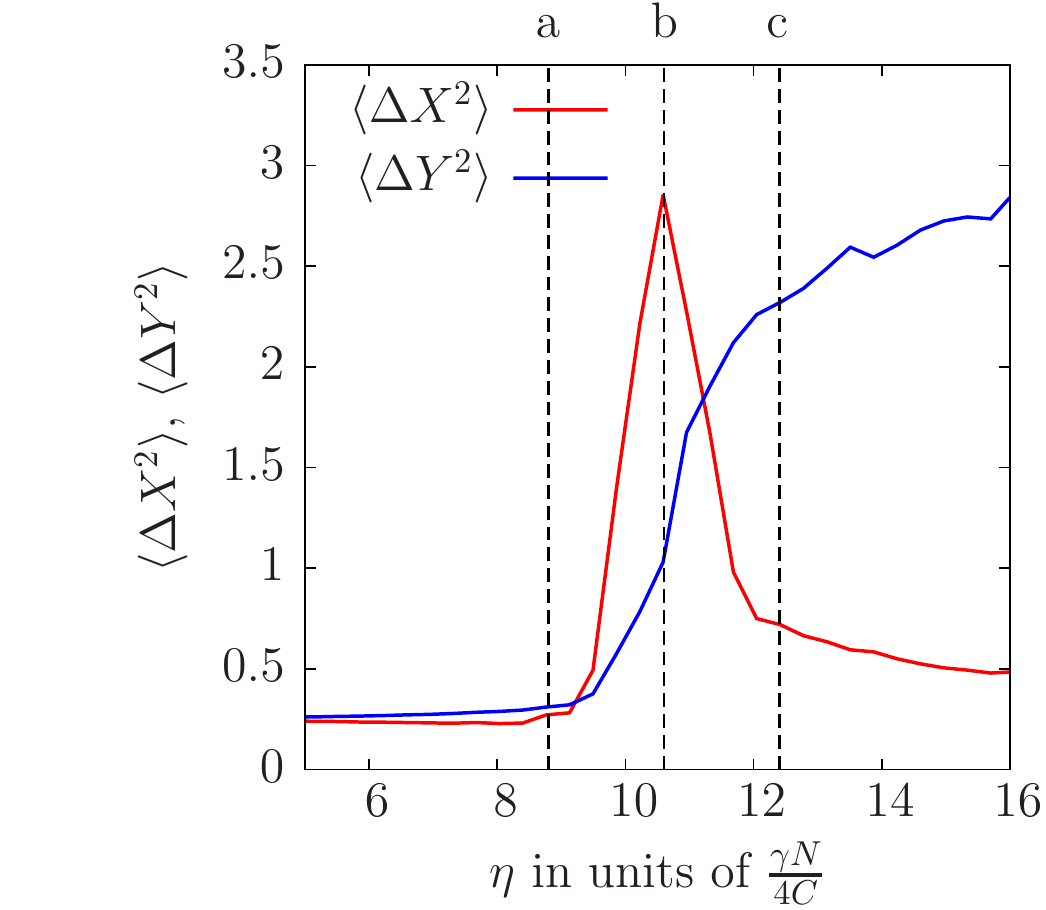}
\end{tabular}
\end{center}
\caption{Wigner function calculated from the cavity steady-state reduced density matrix $\rho_{\rm ss:mode}$ for input amplitude (a) below the bistability regime ($\eta=8.8$), (b) within it ($\eta=10.6$), and (c) above it ($\eta=12.4$). The atom number is $N=6$, the $\eta$ values in question are indicated by vertical dashed lines in \Fref{fig:histogram}. The panel (d) displays the variance of the quadratures $X$ and $Y$ corresponding to the phase space coordinates of the panels (a)-(c).}
\label{fig:Wigner}
\end{figure}

One can associate a Wigner function with the results of the linearized semiclassical model in such a way that it is a Gaussian centred on the mean field values and has a half width corresponding to the variances of the quadratures $\hat X = \frac{a+a^\dagger}{2}$ and $\hat Y= \frac{a-a^\dagger}{2i}$. The Wigner function resulting from the quantum simulations $W\left[\rho_{\rm ss:mode}\right]$ for atom number $N=6$ is plotted in \Fref{fig:Wigner} for various driving strengths. The bimodal distribution corresponding to the double-peaked histogram above can be clearly resolved in the bistability regime as displayed in part (b) of the Figure. The Wigner function is positive everywhere indicating that the two peaks correspond to a mixture of the two bistable possibilities and no coherence between the two markedly different mean fields is present. The quantum state on the lower and upper branches, plots (a) and (c), respectively are significantly different: on the lower branch of the bistability curve the state remains close to a minimal uncertainty coherent state, whereas on the upper branch it develops a banana shape with increased phase uncertainty. 

Figure \ref{fig:Wigner}(d) shows the variance of the $\hat X$ and $\hat Y$ quadratures. The large peak in the variance $\langle\Delta\hat X^2\rangle$ originates,  obviously, from the two-peaked shape of the distribution function, that is, the variance is increased proportionally to the separation of  the two, mixed components. For large driving strength both the $\hat X$ and $\hat Y$ quadratures have larger variances than that of a coherent state ($\langle\Delta\hat X^2\rangle_{\rm coh}=\langle\Delta\hat Y^2\rangle_{\rm coh}=1/4$), especially, the  variance $\langle\Delta\hat Y^2\rangle$ is significantly increased in accordance with the stretched banana shape of the Wigner function.

\begin{figure}[ht]
\begin{center}
\includegraphics[width=0.475\linewidth]{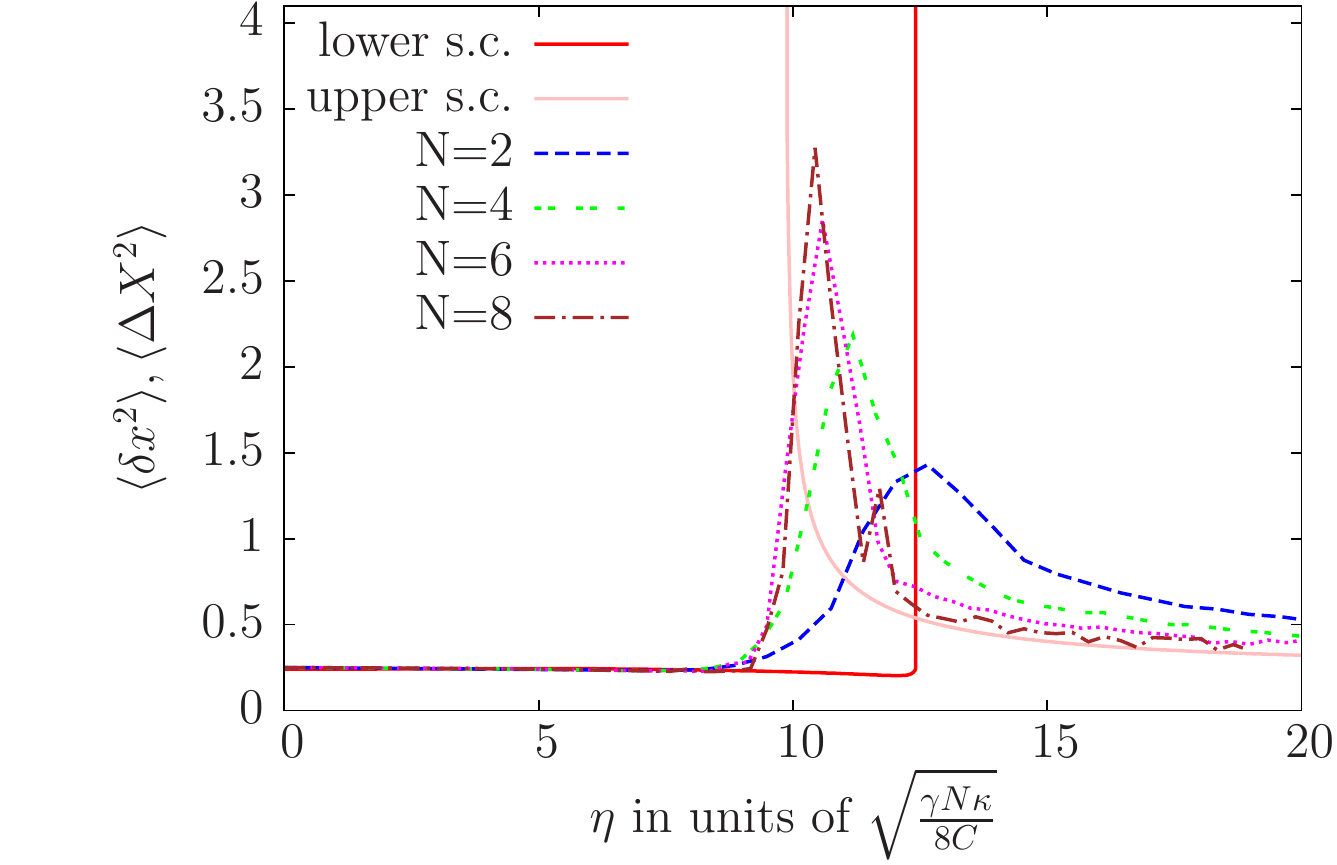}
\includegraphics[width=0.475\linewidth]{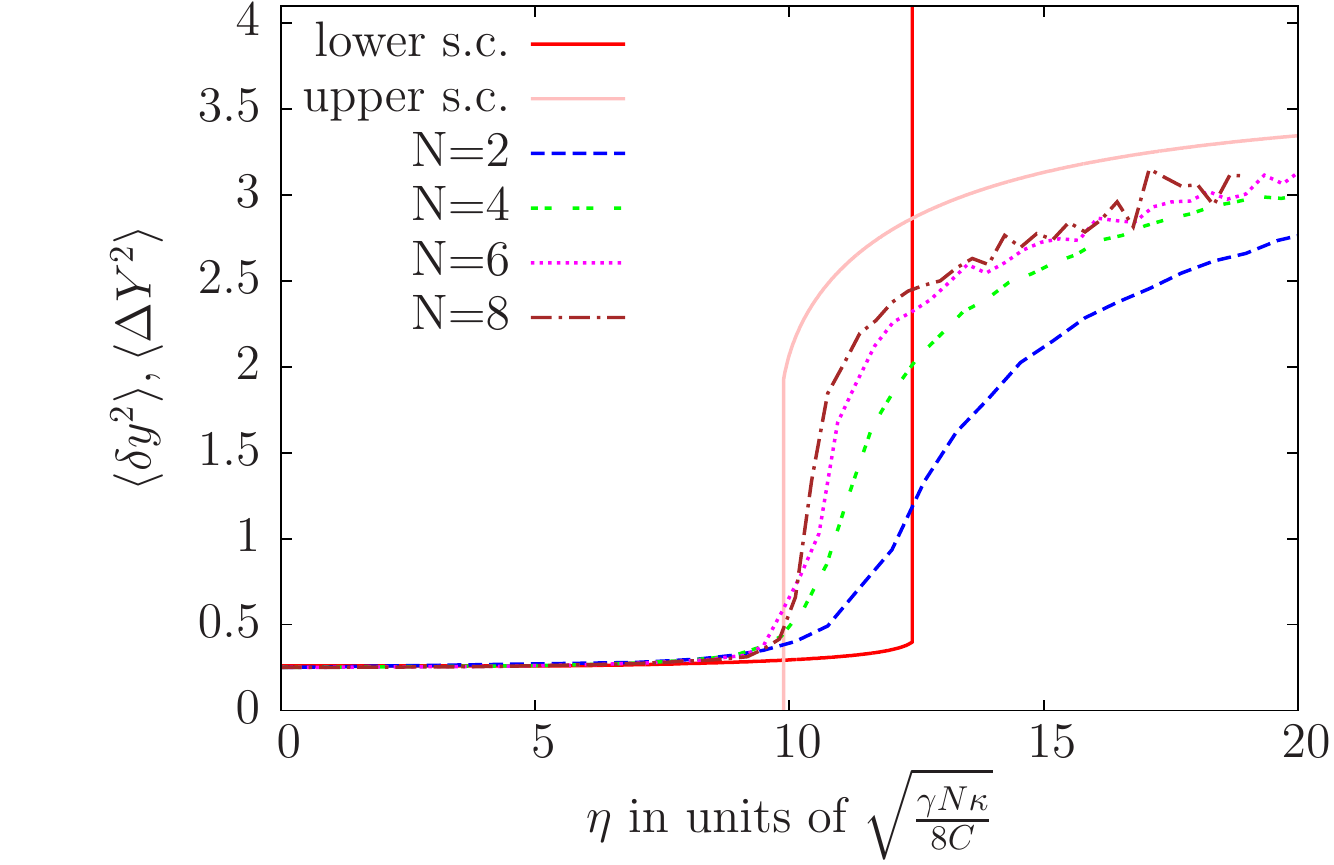}
\caption{Variances $\langle\Delta\hat X^2\rangle$ (left panel) and $\langle\Delta\hat Y^2\rangle$ (right panel) of the $\hat X$ and $\hat Y$ quadratures, respectively, as a function of the drive amplitude. Solid lines represent the first-order autocorrelations of the $\delta x = {\rm Re}\left\{\delta a\right\}$ and $\delta y={\rm Im}\left\{\delta a\right\}$ linearized fluctuations, calculated for $N=4$. They exhibit divergences at the upper and lower boundaries of the lower and upper semiclassical mean field solutions, respectively. These divergences enclose the bistability range where the quadrature variances increase due to the separation of the two peaks mixed within the Wigner function.}
\label{fig:VarianceFiniteSize}
\end{center}
\end{figure}
Figure \ref{fig:VarianceFiniteSize} is used to compare the quadrature variances obtained in the exact quantum model and in the semiclassical approach. In this latter,  the linearized fluctuation analysis of the mean field theory allows for calculating first-order correlation functions. As shown in Fig.~\ref{fig:VarianceFiniteSize} by solid lines, the semiclassical model leads to singularities of the variance at the critical points, i.e., at the boundaries of the bistability region. The divergence appears because the soft mode, one of the eigenmodes of the linear system in Eq.~(\ref{eqs:Linearized}), has a vanishing eigenfrequency at the critical point. This occurs both along the lower and the upper mean-field solution curves. The numerical simulations of the exact quantum model are, of course, exempt from such a singularity.  The increase of the variance in the bistability regimee is not the signature of some finite size regularization of a non-analitic function but, as discussed previously in relation to Figure \ref{fig:Wigner}(d),  it is the manifestation of the bimodal photon number distribution. This latter effect is by definition beyond the scope of the semiclassical approach in which one linearizes around a selected mean field value. 

The quantum calculation exhibits a rapid convergence to the semiclassical results outside the bistability regime, confirming that with a number of atoms N = 8 the common features of optical bistability can be reproduced. On the lower branch of the bistability curve, that is, for small driving strength, the variance renders $\frac{1}{4}$ corresponding to that of a coherent state.  On the upper branch, the variance tends to a value larger than $\frac{1}{4}$ which is obtained for all the quantum calculations performed for different atom numbers. 

\subsection{Atom-light field correlations}

The principal source of nonlinearity in the Jaynes--Cummings model, expressed in the form of Heisenberg--Langevin equations (\ref{eq:HeisenbergLangevin}), is represented by the operator products $a^\dagger \sigma$, $\sigma^\dagger a$,  and $\sigma_z a$.  Within the mean field theory, the mean of these terms are approximated by the product of mean values, e.g., $\sum_i \langle a^\dagger \, \sigma_{i} \rangle \approx  \sum_i \langle a^\dagger \rangle \,\langle \sigma_{i}\rangle $. One can expect such a factorization to hold in a large ensemble, which would validate the semiclassical model. We resorted to this approximation not only in the course of calculating the steady-state mean values but also when neglecting the quadratic noise terms in the linearized fluctuation analysis. As can be seen  for example in Fig.~\ref{fig:steadyStateIntensity}, we obtained a mean field amplitude fitting nicely to the exact quantum results for atom number as low as $N=8$.  Equipped with the quantum Monte Carlo calculation, we can directly check this approximation in the range of atom numbers $N=2\ldots 8$, where we recorded satisfying convergence to the mean field results. 

\begin{figure}
\begin{center}
\includegraphics[width=0.67\linewidth]{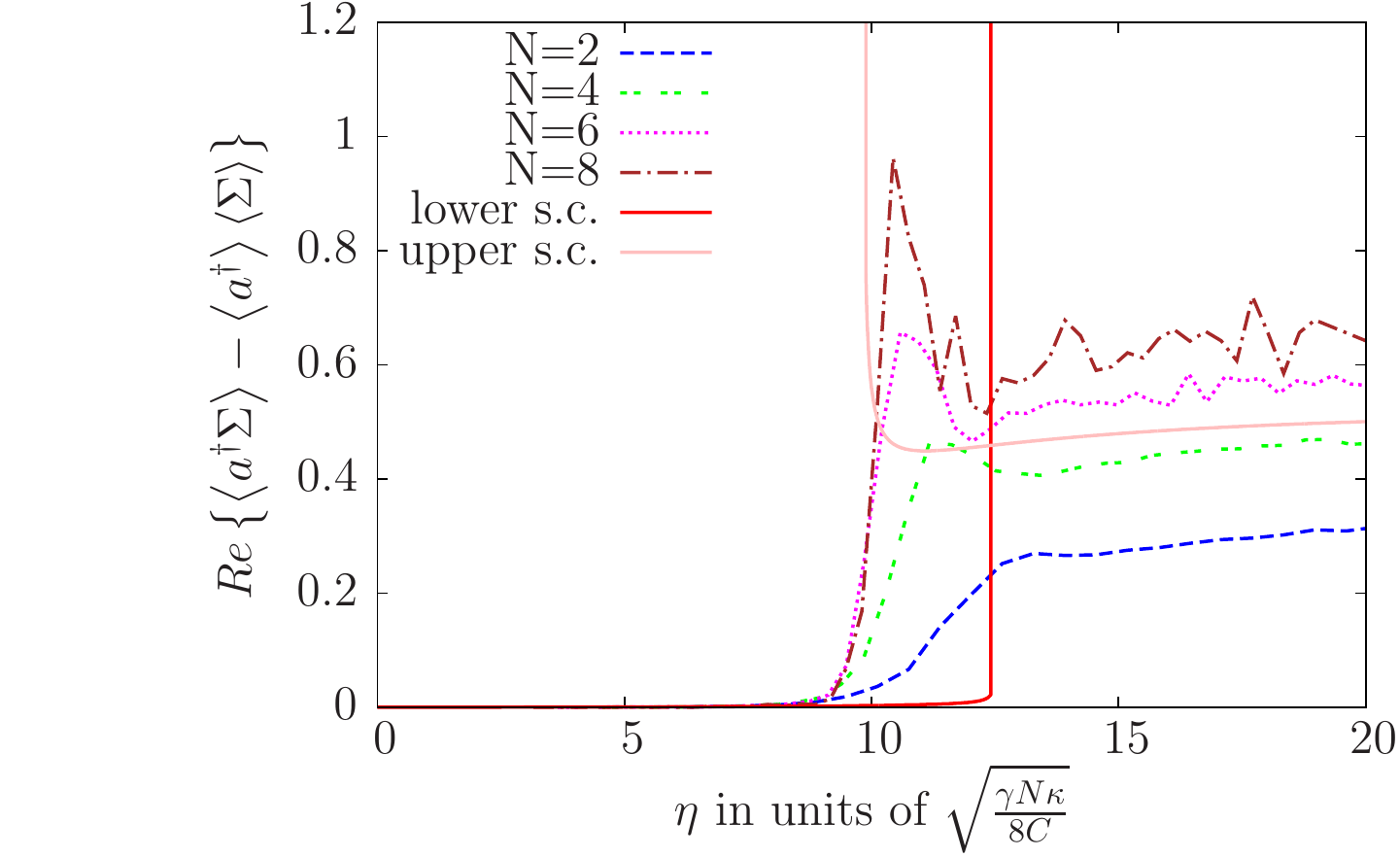}
\caption{The real part of the correlation function $\langle a^\dagger \, \Sigma\rangle - \langle a^\dagger \rangle\, \langle \Sigma\rangle$ as a function of the driving amplitude $\eta$ for atom numbers N= 2, 4, 6, 8. Solid lines show the semiclassical result for linearized quantum fluctuations around the mean field solutions, both the lower and upper mean fields are considered in the bistability range.  }
\label{fig:AtomFieldCorrelation}
\end{center}
\end{figure}
Figure \ref{fig:AtomFieldCorrelation} shows the real part of the first order correlation function $\langle a^\dagger \, \Sigma\rangle - \langle a^\dagger \rangle\, \langle \Sigma\rangle$. For the mean-field model (solid lines), the correlation function is independent of $N$. There is a good agreement between the curves along the lower branch below the bistability range: here the correlation vanishes in all the cases considered. For strong driving, above the bistability domain,  the correlation does not decay but tends to a finite value comparable to that in the bistability range. However, the quantum model reveals a deviation from the semiclassical model: the correlation increases with the atom number and the deviation becomes significant for $N=8$. Other correlation functions exhibit qualitatively similar behaviour. We mention that significant quantum correlations in a strongly driven cavity QED system is not unexpected: similar effect in a micro cavity laser system under strong incoherent pumping have been found between the population inversion and the photon number \cite{Rice1994Photon}.

\section{Conclusion}
\label{sec:Conclusion}

Optical bistability has received considerable interest in recent years owing to its potential for the development of ultra-low power photonic signal processing devices, e.g., optical switches \cite{Yang2011Controllable}. In this paper we showed that the system of a few atoms spatially localised within and strongly coupled to the radiation field of a high-finesse resonator can be operated as a bistable device, albeit in the very low-excitation quantum limit. The semiclassical solutions of absorptive bistability can be well resolved with an atomic medium containing a number of atoms as low as 6 to 8.  Such systems can be created both by using atoms in microwave or optical cavities and by using artificial atoms in circuit QED \cite{Fink2009Dressed}. We showed that the photon statistics and atom-field correlations are qualitatively well described by the linearized fluctuation analysis around the mean field solutions. We find deviations in the strong-driving limit where significant atom-field quantum correlations build up. It remains a subject of further researches if the semiclassical solutions can be stabilized by feedback in the bistability range and if transition between the different branches can be generated deterministically by weak external modulation of the driving amplitude. 


\bibliographystyle{unsrt}
\bibliography{optical_bistability}

\end{document}